\documentclass[twocolumn,prl,showpacs,superscriptaddress]{revtex4-1}
\usepackage{graphicx}
\usepackage{rotating}
\usepackage{amsmath}
\usepackage{amsfonts}
\usepackage{amssymb}
\usepackage{enumerate}
\usepackage{longtable}
\usepackage{units}
\usepackage{hyperref} 
\usepackage[T1]{fontenc}
\usepackage[utf8]{inputenc}
\usepackage{graphicx}
\usepackage{capt-of}
\usepackage{bbm}
\usepackage{dcolumn}
\usepackage{bm}
\usepackage{color}

\usepackage{epstopdf}
\usepackage{siunitx}
\usepackage{booktabs}
\usepackage{braket}

\begin{document}
\newcommand{\cred}{\color{red}}
\title{Pressure dependence of the magnetization plateaus of SrCu$_2$(BO$_3$)$_2$}
  \author{David A. Schneider}
 \affiliation{Lehrstuhl f\"ur Theoretische Physik I, Otto-Hahn-Str.~4, TU Dortmund, D-44221 Dortmund, Germany}
 
 \author{Kris Coester}
 \affiliation{Lehrstuhl f\"ur Theoretische Physik I, Otto-Hahn-Str.~4, TU Dortmund, D-44221 Dortmund, Germany}

 \author{Fr\'ed\'eric Mila}
 \affiliation{Institut de théorie des phénomènes physiques, École Polytechnique Fédérale de Lausanne (EPFL), CH-1015 Lausanne, Switzerland}

 \author{Kai Phillip Schmidt}
 \affiliation{Lehrstuhl f\"ur Theoretische Physik I, Otto-Hahn-Str.~4, TU Dortmund, D-44221 Dortmund, Germany}

\date{\rm\today}

\begin{abstract}
We show that the critical fields of the magnetization plateaus of the Shastry-Sutherland model decrease significantly upon increasing the ratio of
inter- to intra-dimer
coupling, and accordingly that the magnetization plateaus of SrCu$_2$(BO$_3$)$_2$ shift to lower field under pressure, making the first two
plateaus at $1/8$ and $2/15$ potentially accessible to neutron scattering experiments. These conclusions are based on the derivation of an
effective classical model of interacting pinwheel-shaped spin-2 bound states using a combination of perturbative and graph-based continuous unitary transformations, showing that pinwheel crystals are indeed the lowest-energy plateau structures at low magnetization, and that a simple model of 
intermediate-range two-body repulsion between pinwheels is able to account quantitatively for the plateau sequence. 
\end{abstract}


\maketitle

{\it Introduction--}
\label{Sect:Intro}
The frustrated quantum magnet SrCu$_2$(BO$_3$)$_2$ is one of the most important players in the field
 of highly frustrated quantum magnetism \cite{bookquantummagnetism} due to its very rich and
 complex magnetization curve \cite{Kageyama99,Onizuka00,Kageyama00,Kodama02,Takigawa04,Levy08,Sebastian08,Jaime12,Takigawa13,Matsuda13}. 
Indeed, experiments in ultrastrong magnetic fields unveil a multitude of intriguing behaviors such as a series of magnetization plateaus archetypical of frustrated quantum magnetism.  Despite a huge body of literature over the last 15 years \cite{Kageyama99,Onizuka00,Kageyama00,Kodama02,Takigawa04,Levy08,Sebastian08,Jaime12,Takigawa13,Matsuda13,Miyahara99,Momoi00a,Momoi00b,Fukumoto00,Fukumoto01,Miyahara03a,Miyahara03b,Dorier08,Abendschein08,Nemec12,Lou12,Corboz14}, a consistent theoretical understanding of the full sequence of plateaus seemingly emerged only recently \cite{Corboz14}. The low part of the magnetization curve is most exciting since the magnetization plateaus are predicted to be built from exotic pinwheel-shaped spin-2 bound states \cite{Corboz14} and not from individual triplon excitations \cite{Schmidt03}.  

Triplons are indeed the natural building block for the magnetization plateaus in SrCu$_2$(BO$_3$)$_2$, since the underlying microscopic description corresponds remarkably well to the quantum $S=1/2$ Heisenberg antiferromagnet on the two-dimensional Shastry-Sutherland lattice. The Shastry-Sutherland model \cite{Shastry81} can be seen as a set of mutually orthogonal dimers which are coupled by an interdimer coupling $J'$. Its beauty arises from the exact ground state in terms of a product state of singlets at zero magnetic field. One natural approach is then to view the magnetization process as populating the dimers by triplets. Even though this approach gives important insights for the plateaus at 1/3 and 1/2 \cite{Dorier08,Nemec12,Matsuda13}, the prediction in terms of crystals of triplons for the plateau sequence at low magnetization disagrees with the experimental one, whereas pinwheels naturally give rise to the experimental sequence $1/8$, $2/15$, and $1/6$ \cite{Corboz14}.

Neutron scattering of the low-magnetization plateaus $1/8$, $2/15$, and $1/6$ represents a powerful tool to clarify the magnetization structure of these plateaus, but experiments are challenging due to the rather large external magnetic fields needed (27 T for the first plateau at 1/8). However, external pressure on SrCu$_2$(BO$_3$)$_2$ is known to reduce the exchange couplings $J$ and $J'$ and to increase the ratio $J'/J$ \cite{Zayed2010}, which implies that the critical
fields of the plateaus will change with pressure. Consequently, it is of major importance to predict the sequence of plateaus as a function of field and pressure, and to calculate the experimental signatures of pinwheel crystals in neutron scattering experiments.        

This is exactly the objective of this Letter. First, we establish an effective low-energy model in terms of interacting pinwheels which captures the low-magnetization plateaus of the frustrated quantum magnet SrCu$_2$(BO$_3$)$_2$ quantitatively. The flexibility of our approach allows us to predict the evolution of pinwheel crystals when varying microscopic coupling constants and to show  that the critical magnetic fields of the low-magnetization plateaus decrease with pressure, making the pinwheel crystals at $1/8$ and $2/15$ accessible to neutron scattering with experimentally realistic values of field and pressure. In addition, we calculate the local magnetization of these plateaus which is of direct importance for elastic neutron scattering experiments on SrCu$_2$(BO$_3$)$_2$.

\begin{figure}[t]
\centering
\includegraphics[width=\columnwidth]{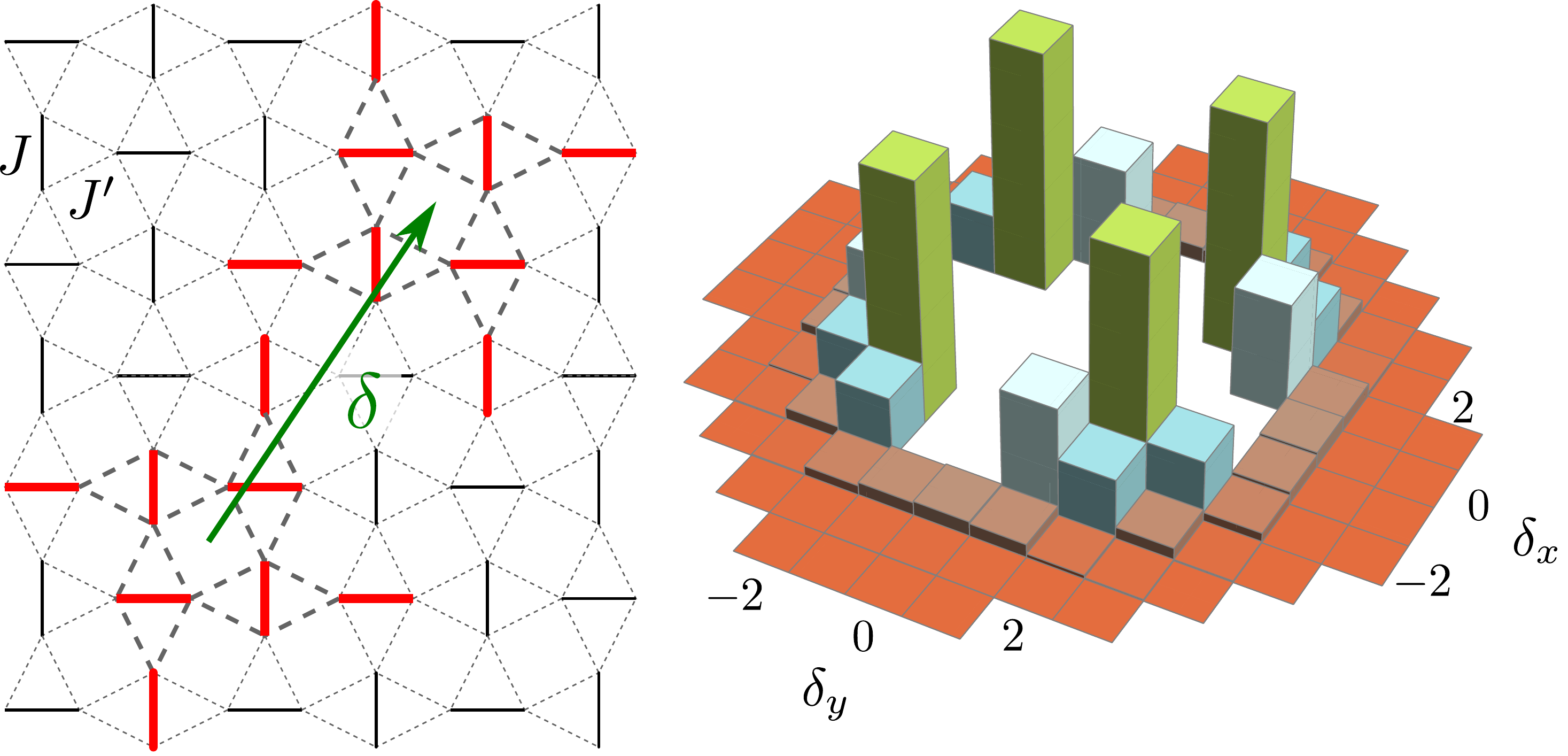}
\caption{{\it Left}: Illustration of the Shastry-Sutherland model. Solid (dashed) bonds denote intra-dimer (inter-dimer) couplings $J$ ($J'$). Additionally, two pinwheels at distance \mbox{$\mathbf{\delta}=(\delta_x,\delta_y)$} are shown as thicker red bonds. {\it Right}: The effective interaction $V_{\rm pw}(\mathbf{\delta})$ between two pinwheels located at sites $(0,0)$ and $\mathbf{\delta}$ of the effective square lattice formed by the centers of the pinwheels for $J'/J=0.63$. Note that the inner hardcore potential for small $\mathbf{\delta}$ is not given (white area).}
\label{pic:model}
\end{figure}

{\it Model--}
\label{Sect:model}
We study the spin-$1/2$ Shastry-Sutherland model in an external magnetic field $h$ which reads
\begin{equation}
\mathcal{H} = J \sum\limits_{\langle i,j \rangle} \mathbf{S}_i \cdot \mathbf{S}_j  + J' \sum\limits_{\langle\langle i,j \rangle\rangle} \mathbf{S}_{i} \cdot \mathbf{S}_{j} - h \sum\limits_i S^z_i \quad ,
\label{eq:ham}
\end{equation}
with the bonds  $\langle i,j\rangle$  building an array of orthogonal dimers and the bonds  $\langle\langle i,j \rangle\rangle$ representing inter-dimer couplings. The geometry of the Shastry-Sutherland model and its two-dimer unit cell, made of a vertical and a horizontal dimer, is illustrated in Fig.~\ref{pic:model}.

Here we focus on the phase diagram of Eq.~\eqref{eq:ham} for parameter ratios $J'/J\le 0.675$ for which the pure Shasty-Sutherland model is in the exact product state of singlets for $h=0$ \cite{Lou12,Corboz13}, since this is the relevant coupling regime for SrCu$_2$(BO$_3$)$_2$ \cite{Matsuda13}.
%
%

{\it Approach--}
\label{Sect:approach}
We apply a continuous unitary transformation (CUT) to map the Shastry-Sutherland model onto an effective low-energy model describing triplons on a square lattice along the lines of Ref.~\onlinecite{Dorier08}. The essential feature of the derived effective Hamiltonian is the conservation of the total number of triplons. In a finite magnetic field, the relevant triplon states have maximal $S_z$. Other channels are only important if bound states of triplons with different quantum numbers become relevant at low energies \cite{Manmana11}. Furthermore, we concentrate on the plateaus at low densities where pinwheels are expected. Pinwheels correspond to two-triplon bound states with $S=2$ stabilized by two-triplon correlated hopping processes \cite{Momoi00b,Foltin14}. We therefore restrict the terms in the effective model to one- and two-particle operators
\begin{eqnarray}
\mathcal{H}_\text{eff} &=& \mu\sum_i \hat{n}_i  + \sum_{i,j} V_{ij}\, \hat{n}_i \hat{n}_j + \sum_{i,j,k} t^{ijk}_{\rm c}\left(  b^\dagger_i b_j^{\phantom{\dagger}} \hat{n}_k^{\phantom{\dagger}} + {\rm h.c.}\right) \nonumber\\
&&+ \sum_{i,j,k,l} P_{ijkl}\, b^\dagger_i b^\dagger_j b_k^{\phantom{\dagger}} b_l^{\phantom{\dagger}} \quad ,
\end{eqnarray}
where the sums run over the effective square lattice built by the $J$-dimers (see also Fig.~\ref{pic:model}). The hardcore boson operator $b_i^\dagger$ ($b_i^{\phantom{\dagger}}$) creates (annihilates) a triplet $|t^1\rangle$ at site $i$, and $\hat{n}_i=b^\dagger_i b_i^{\phantom{\dagger}}$ is the 
local density operator with eigenvalues $0$ and $1$.

In Ref.~\onlinecite{Dorier08}, the coefficients of $\mathcal{H}_\text{eff} $ have been determined perturbatively up to high orders in $J'/J$ applying perturbative CUTs (pCUTs) \cite{Knetter00,Knetter03}. This expansion gives a satisfactory effective model up to $J'/J\approx 0.5$. For larger values of $J'/J$, while extrapolations of the chemical potential $\mu$ and of the two-particle interactions $V_{ij}$ still work satisfactorily, no consistent extrapolation of correlated hopping processes $t^{ijk}_{\rm c}$ exists.

As a consequence, we use the following strategy, which is well adapted to the physics of pinwheel crystals: i)  We apply non-perturbative graph-based CUTs (gCUTs) \cite{Yang11,Coester15} to treat the rather local quantum fluctuations like correlated hopping. ii) We take the longer-range part of the (extrapolated) two-triplon repulsive density-density interactions $V_{ij}$ obtained by pCUTs to quantitfy the interaction between pinwheels. This is reasonable, since the entanglement between different pinwheels is expected to be low. Let us stress that the {\it same} quasi-particle generator for the CUT is used in pCUTs and gCUTs, so it is indeed the same CUT which is performed for all couplings of the effective model. The effective Hamiltonian takes then the form \mbox{$\mathcal{H}_\text{eff} = \mathcal{H}_{\rm eff}^{\rm gCUT} + \mathcal{H}_{\rm eff}^{\rm pCUT}$}. The chemical potential $\mu$ and (most of) the repulsive interactions $V_{ij} $ are taken from Ref.~\onlinecite{Dorier08}.  All other two-particle processes are deduced by gCUTs and are contained in $\mathcal{H}_{\rm eff}^{\rm gCUT}$.

The general idea of gCUTs \cite{Yang11,Coester15} is to perform an exact CUT on topologically distinct graphs. Here we do not perform a full graph expansion, but we consider the single C$_4$-symmetric cluster of $8$ dimers as shown in Fig.~\ref{pic:model}, since this is the smallest cluster on which a pinwheel fits. From the cluster-dependent amplitudes we extract the closest to the thermodynamic limit approximations of the two-triplon amplitudes \cite{Supp}. In contrast to pCUT calculcations where all quantum fluctuations up to a certain order in $J'/J$ are included, the non-perturbative amplitudes contain all processes to infinite order in $J'/J$ fitting on the graphs under consideration. This renders the non-perturbative analysis more powerful in many cases.

The Hamiltonian $\mathcal{H}_{\rm eff}$ represents an interacting hardcore boson model with exotic kinetic terms. It can describe conventional crystals made of single triplons as well as pinwheel crystals. The conventional plateaus of triplons are well treated in the classical limit along the lines of Ref.~\onlinecite{Dorier08}, which amounts to looking for product wave-functions of localized triplons. A quantitative description of pinwheels is obtained by the following procedure. We assume that different pinwheels only interact via the diagonal interactions $V_{ij}$, i.e.~pinwheel crystals are well described by product wave functions of single pinwheels living on the $8$-dimer cluster shown in Fig.~\ref{pic:model}. The centers of pinwheels build again an effective square lattice. We therefore solve the two-particle problem on this cluster using $\mathcal{H}_\text{eff}$ which gives access to the chemical potential $\mu_{\rm pw}$ of a pinwheel as well as its density profile $n_i=\langle \hat{n}_i\rangle$ for all dimers $i$ of a given pinwheel. The latter allows to determine the effective interaction $V_{\rm pw}(\mathbf{\delta})$ between two pinwheels at distance $\mathbf{\delta}$ (see Fig.~\ref{pic:model} right) by summing up the various two-particle interactions $V_{ij} n_i n_j$ so that $i$ and $j$ correspond to dimers of different pinwheels. One then obtains the following effective pinwheel Hamiltonian
\begin{equation}
 \mathcal{H}_{\rm pw}=\mu_{\rm pw}\sum_{\mathbf{i}} \hat{n}^{\rm pw}_{\mathbf{i}} + \sum_{\mathbf{i},\mathbf{\delta}}  V_{\rm pw}(\mathbf{\delta})\, \hat{n}^{\rm pw}_{\mathbf{i}} \hat{n}^{\rm pw}_{\mathbf{i}+\mathbf{\delta}}\quad ,
 \label{eq:pwham}
\end{equation}
where $\mathbf{i}$ runs over the sites of the effective square lattice built by the centers of pinwheels and $\hat{n}^{\rm pw}_{\mathbf{i}}$ is the local density operator of pinwheels with eigenvalues $0$ and $1$. This procedure can be further optimized by determining the local particle densities $n_i$ inside the pinwheels self-consistently (see Ref.~\onlinecite{Supp}). However, all essential properties discussed below are already present without this self-consistency loop.

The effective model $\mathcal{H}_{\rm pw}$ is essentially classical since it only contains pinwheel density operators. This implies that the treatment of pinwheel crystals becomes extremely simple and transparent. In particular, the energy can be extracted by summing up the relevant pinwheel interactions $V_{\rm pw}(\mathbf{\delta})$. One can also calculate the local magnetizations \mbox{$m_i=\langle S_i^z \rangle$} of any {\it spin} site of the Shastry-Sutherland model using the same kind of approximation. To this end we transform the observables $S_i^z$ for each of the $16$ spin sites on a single pinwheel cluster by the same CUT  and calculate the expectation values with respect to the ground-state solution of the effective model $\mathcal{H}_\text{eff}$.

\begin{figure}[t]
\centering
\includegraphics[width=\columnwidth]{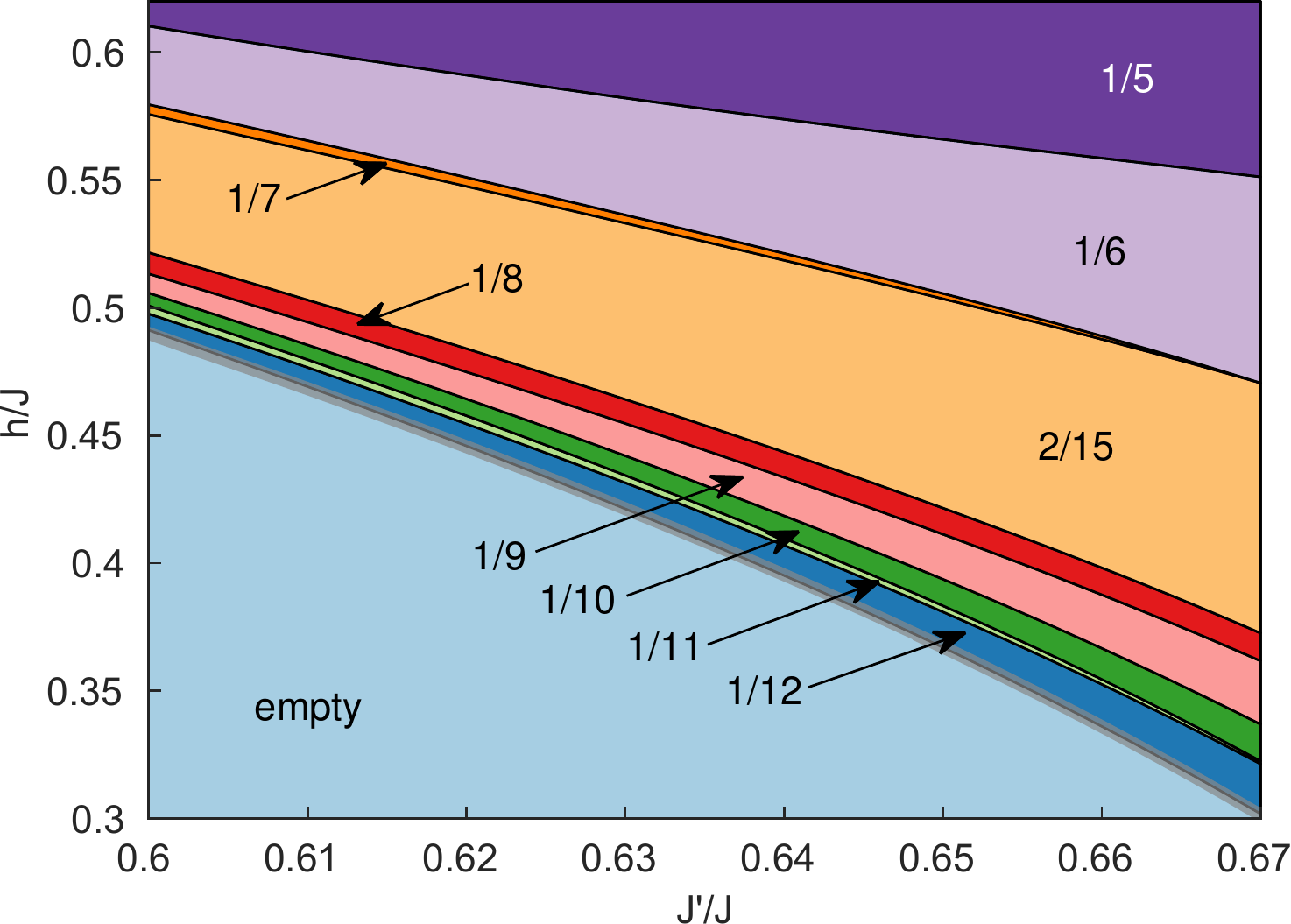}
\begin{center}
\includegraphics[width=0.44\columnwidth]{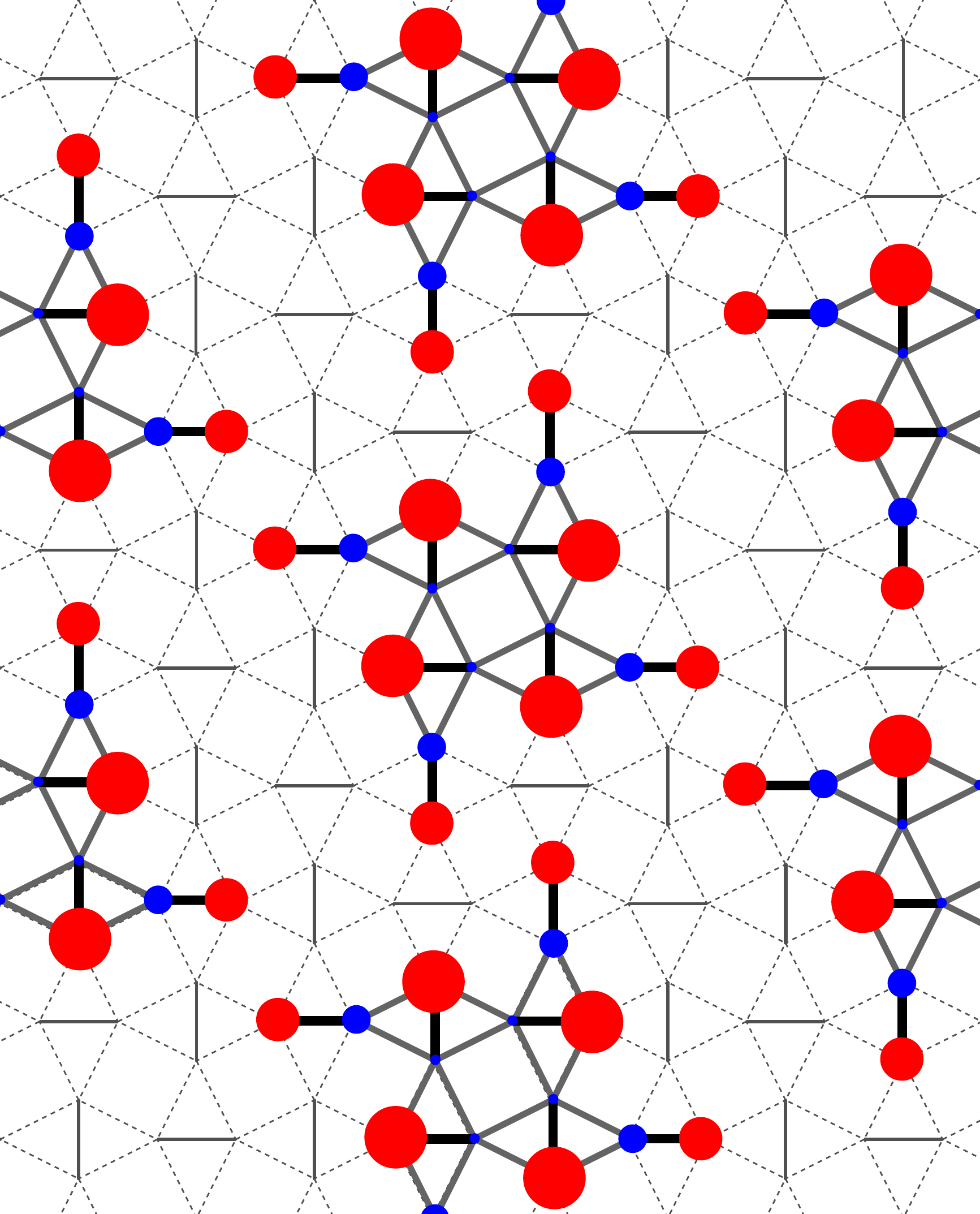}\hspace*{5mm}
\includegraphics[width=0.44\columnwidth]{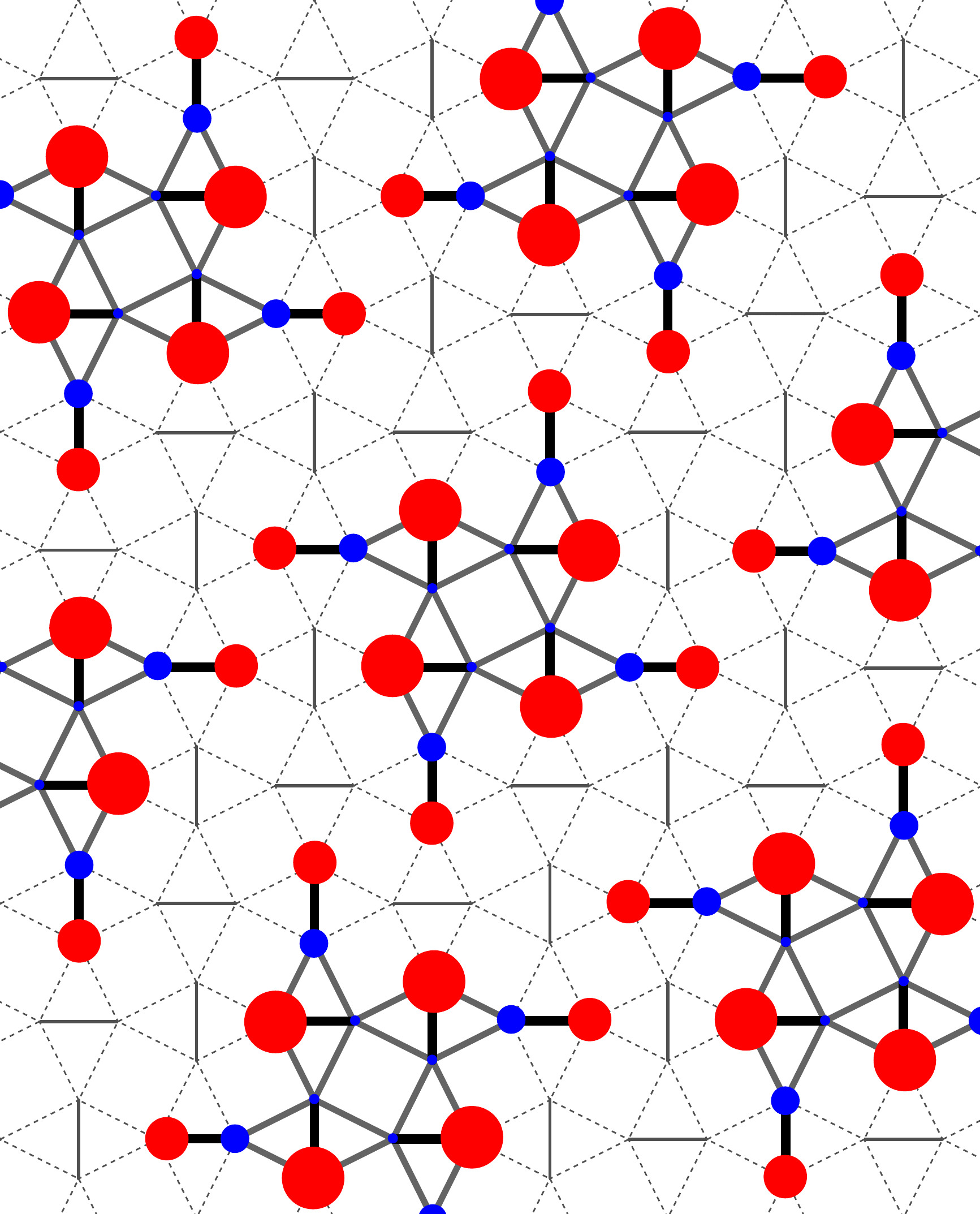}
\end{center}
\caption{{\it Upper panel}: Phase diagram of the Shastry-Sutherland model as a function of $J'/J$ and magnetic field $h/J$ obtained from the effective pinwheel model~$\mathcal{H}_{\rm pw}$ in a self-consistent calculation. {\it Lower panel}: Local magnetization of the pinwheel crystals at densities 1/8 (left) and 2/15 (right). The area of the red (blue) circles is proportional to the positive (negative) magnetization along the $z$-direction orthogonal to the displayed plane.}
\label{pic:pd}
\end{figure}

{\it Phase diagram --} We have calculated the energy of both types of crystal structures, along the lines of Ref.~\onlinecite{Dorier08} for the conventional 
triplon crystals, and by solving the effective pinwheel model of Eq.~\eqref{eq:pwham} for the pinwheel crystals. Remarkably, the energy of conventional plateaus is considerably larger than that of the competing pinwheel crystals for $M\le 1/4$ and all values of $J'/J$, resulting in the phase diagram displayed in Fig.~\ref{pic:pd} that contains only pinwheel crystals. The results are in excellent agreement with recent numerical calculations \cite{Corboz14,Footnote1}.
In particular, it contains the experimental sequence $1/8$, $2/15$, and $1/6$, and the arrangements of pinwheels are identical. 
The same is even true for the tiny domain-wall crystal at density 1/7 which is located between 2/15 and 1/6 \cite{Footnote2}. Furthermore, our energies of the various plateaus are in quantitative agreement with those of Ref.~\onlinecite{Corboz14} for $J'/J=0.63$, the largest deviation between both calculations being below $10^{-3}J$ and only $\approx 10^{-4}J$ for 1/8 and 2/15. Note that our phase diagram also contains lower
magnetization plateaus which are not present in experiments, presumably because the pinwheels can delocalize at very low density, a possibility not included
in our model which is aimed at comparing plateau structures among themselves but not with compressible phases.

The effective model $\mathcal{H}_{\rm pw}$ therefore represents a quantitative, light, and physically intuitive description of pinwheel crystals in the Shastry-Sutherland model in a broad range of ratios $J'/J$ including the relevant regime for the frustrated quantum magnet SrCu$_2$(BO$_3$)$_2$. These properties of $\mathcal{H}_{\rm pw}$ are exploited in the following to give precise predictions in order to identify pinwheel crystals experimentally. 
%
%

{\it SrCu$_2$(BO$_3$)$_2$ under pressure --} One may wonder how to unambiguously identify pinwheel crystals experimentally keeping in mind the rather large critical fields of the compound SrCu$_2$(BO$_3$)$_2$, e.g.~$27$\,T for the lowest plateau at 1/8. The best option to pinpoint pinwheel crystals is certainly neutron scattering which is however challenging at these large field values. Here we propose to apply external pressure turning neutron scattering experiments on the pinwheel crystals of SrCu$_2$(BO$_3$)$_2$ into a realistic and valuable option.

\begin{figure}[t]
\centering
\includegraphics[width=\columnwidth]{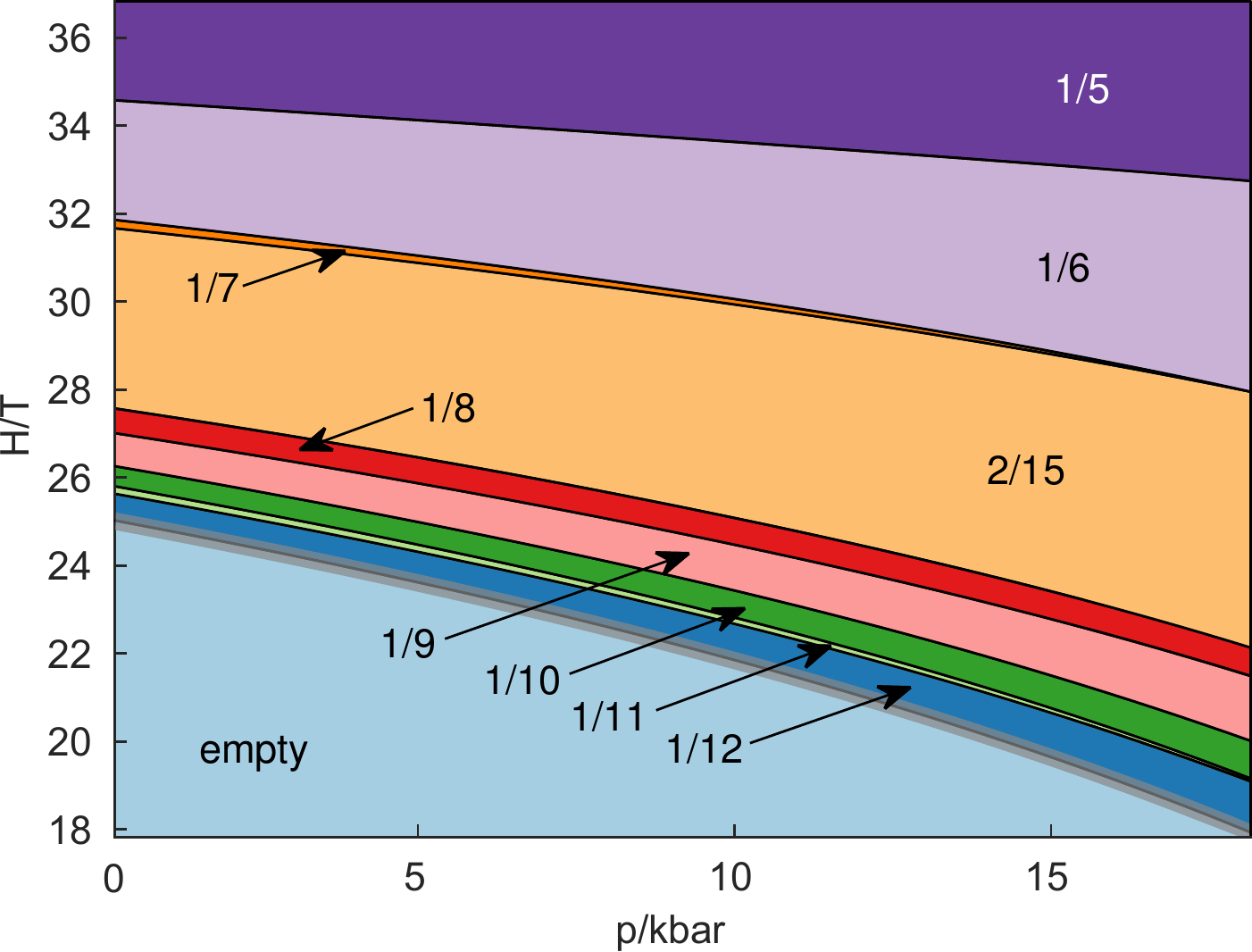}
\caption{Phase diagram of SrCu$_2$(BO$_3$)$_2$ as a function of pressure $p$ and magnetic field $H$ obtained from the effective pinwheel model~$\mathcal{H}_{\rm pw}$.}
\label{pic:pd_pressure}
\end{figure}

The effect of pressure $p$ is known from zero-field measurements of the magnetic susceptibility to be twofold in SrCu$_2$(BO$_3$)$_2$ \cite{Zayed2010}. First, the absolute values of $J$ and $J'$ are reduced. Second, the ratio $J'/J$ is further increased, but one stays in the same zero-field phase as long as \mbox{$p\leq 17$\,kbar}, which defines the relevant pressure window. To a good approximation \cite{Zayed2010}, the coupling constants $J$ ($J'$) decrease linearly with a slope $-0.63 \ \text {T/kbar}$ ($-0.29 \ \text {T/kbar}$) enabling us to convert our theoretical phase diagram Fig.~\ref{pic:pd} as a function of $J'/J$ into one where the critical magnetic fields of the various plateaus is shown as a function of pressure $p$. We fix the zero-pressure values to \mbox{$J=59.4$\,T} and \mbox{$J'=37.4$\,T} ($J'/J\approx 0.63$) so that we recover the critical field $H_c=27$\,T for the lowest 1/8 plateau. The resulting phase diagram is displayed in Fig.~\ref{pic:pd_pressure}.
%
%

As expected, all critical fields for the various plateaus decrease with increasing pressure. 
Experimentally, neutron scattering can be done up to fields of $26$\,T with nowadays technology, i.e.~one almost reaches the lowest 1/8 plateau ($H_c=27$\,T at ambient pressure). Interestingly, already a moderate pressure between $5$\,kbar and $10$\,kbar is sufficient to push the critical field of the 1/8 plateau below $26$\,T. Furthermore, one will also access the 2/15 plateau when applying a pressure of $10$\,kbar. These results therefore clearly demonstrate that neutron scattering at realistic values of the magnetic field and at moderate pressure is a promising option to determine the structure of the lowest two magnetization plateaus of the frustrated quantum magnet SrCu$_2$(BO$_3$)$_2$, and to identify pinwheels as the elementary building blocks of these exotic states of quantum matter. Indeed, the Bragg peaks will give access to the unit cells (which are specific to pinwheel crystals as opposed to triplon crystals), and the form factor to important information on the local magnetization inside the unit cell.

Finally,  it is known from NMR measurements that the widths of the 1/8 and 2/15 plateaus are similar and of the order $1$\,T, and that an incommensurate phase exists between the plateaus at 2/15 and 1/6 in which the translational symmetry is still broken while the magnetization raises monotonically with field \cite{Takigawa13}. If one assumes that the width of the 2/15 plateau remains small when applying moderate pressure, our results also predict that this phase should be accessible by future neutron scattering experiments.

{\it Discussion--}
\label{Sect:conclusion}
We have derived an effective low-energy model directly in terms of pinwheels by. The quantitative agreement in the experimentally relevant coupling regime $J'/J=0.63$ between our results and the 
numerical ones using tensor network calculations \cite{Corboz14} implies that pinwheel crystals can be very well
understood as product-wave functions where the elementary building blocks are individual pinwheels 
living on eight dimers of the Shastry-Sutherland lattice. Each pinwheel corresponds microscopically to a two-triplon 
bound state stabilized by correlated hopping processes. 

The effective pinwheel model $\mathcal{H}_{\rm pw}$ is purely classical since
 only the location of pinwheels matters. In fact, one can rewrite this effective Hamiltonian exactly as an 
antiferromagnetic Ising model on the square lattice with peculiar types of geometrically frustrated 
Ising interactions quantified by $V_{\rm pw}(\delta )$. In this picture, it is the competition between these frustrated interactions 
that leads to a sequence of pinwheel crystals that includes the experimentally relevant ones at 1/8, 2/15, and 1/6. Let us remark that the 1/5 plateau is expected to be unstable under the inclusion of DM-interactions \cite{Corboz14}. 

In addition, this effective pinwheel model potentially offers a natural explanation for the incommensurate regime between the plateaus at 2/15 and 1/6 reported in NMR experiments in terms of a devil's staircase and the associated proliferation of domain walls, a phenomenon well identified in the phase diagram of frustrated Ising models \cite{Bak,Aubry}. Unfortunately, we have found no devil's staircase in the effective model $\mathcal{H}_{\rm pw}$. The stabilization of such a devil's staircase might require additional ingredients such as DM-interactions, known to be present in SrCu$_2$(BO$_3$)$_2$, or longer-range pinwheel interactions. This is an interesting subject for future investigation.

To summarize, we have found a physically intuitive and quantitative description of pinwheel crystals in the frustrated quantum magnet SrCu$_2$(BO$_3$)$_2$. This has allowed us to come up with detailed predictions for future experiments like the evolution of pinwheel crystals under pressure. In future investigations it would be interesting to see whether one can also calculate the dynamical structure factors of pinwheel crystals, and whether superfluid or supersolid phases can also be addressed in our framework by melting the observed pinwheel crystals. 

\acknowledgments
We thank P. Corboz, B. Lake, H. Ronnow and M. Zayed for fruitful discussions. This work was supported by the Helmholtz Virtual Institute 
``New states of matter and their excitations'' as well as from the Deutsche Forschungsgemeinschaft (DFG) with grant SCHM 2511/9-1, and by the Swiss National Science Foundation.

\pagebreak

\begin{widetext}
\begin{center}
\textbf{\large Supplemental Material}
\end{center}
\end{widetext}
\setcounter{equation}{0}
\setcounter{figure}{0}
\setcounter{table}{0}
\setcounter{page}{1}
\makeatletter
\renewcommand{\theequation}{S\arabic{equation}}
\renewcommand{\thefigure}{S\arabic{figure}}
\renewcommand{\bibnumfmt}[1]{[S#1]}
\renewcommand{\citenumfont}[1]{S#1}
\section{Solution of the two-triplon problem}

In this section, we give some technical details on the procedure of deriving and solving the effective low-energy model which captures the physics of the two-triplon problem.

The relevant part of the effective low-energy model describing triplons on a square lattice is given by
\begin{eqnarray}
\mathcal{H}_\text{eff} &=& \mu\sum_i \hat{n}_i  + \sum_{i,j} V_{ij}\, \hat{n}_i \hat{n}_j + \sum_{i,j,k} t^{ijk}_{\rm c}\left(  b^\dagger_i b_j^{\phantom{\dagger}} \hat{n}_k^{\phantom{\dagger}} + {\rm h.c.}\right) \nonumber\\
&&+ \sum_{i,j,k,l} P_{ijkl}\, b^\dagger_i b^\dagger_j b_k^{\phantom{\dagger}} b_l^{\phantom{\dagger}}
\label{eq:heffsup}
\end{eqnarray}
as described in the main text. In the thermodynamic limit, the appearing processes are translationally invariant and can be classified according to their graphical structure. The main task in the derivation of the effective model is to obtain good approximations of these amplitudes, which can be done either in the form of (translationally invariant) series expansions up to a certain order in $J'/J$ (pCUTs) or non-perturbatively (i.e. containing contributions up to infinite order in $J'/J$) on finite clusters of a certain size (gCUTs).

\begin{figure}
\includegraphics[width=0.4\columnwidth]{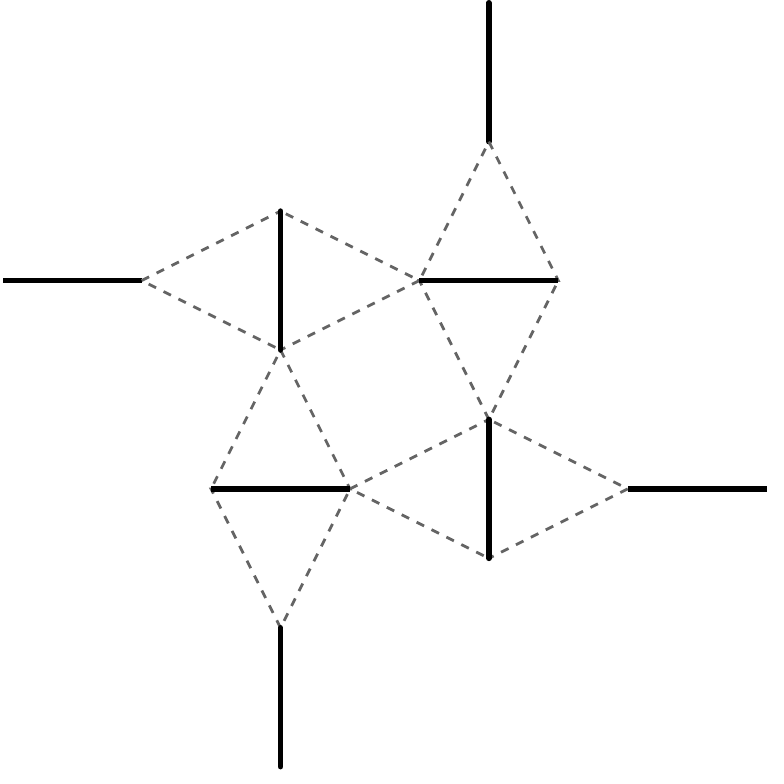}
\hspace{-5pt}
\resizebox{0.55\columnwidth}{!}{
\input{tdens_vs_jprime}
}
\caption{\emph{Left panel}: Finite cluster for the analysis of the bound state of two triplets. \emph{Right panel}: Inner triplon density (ITD) as a function of $J'/J$ as obtained by solving the effective model on the finite cluster shown on the left.}
\label{fig:cluster8}
\end{figure}

While the diagonal amplitudes $\mu$ and the interactions $V_{ij}$ are available in the form of well-converged series expansions, the amplitudes corresponding to the two dominant classes of kinetic processes (correlated hopping $t^{ijk}_{\rm c}$ and pair hopping $P_{ijkl}$) as well as the two next-neighbour (i.e. close-range) interactions are very hard to extrapolate within the experimentally accessible regime of $J'/J$. As the pinwheel bound state of two triplons is strongly kinetically driven, these processes are expected to be of major importance for their quantitative analysis. To overcome this problem, we apply the gCUT scheme to the $C_4$-symmetric cluster shown in Fig. \ref{fig:cluster8}. 

This includes as a first step an exact CUT on this cluster. The resulting Hamiltonian takes the same form as in Eq.\ \eqref{eq:heffsup} with two differences: First, all summing indices are restricted to the finite cluster. Second, the obtained amplitudes are obviously not translationally invariant. Consider for example the chemical potential; it is larger on the outer dimers than on the inner ones due to the difference in the number of available fluctuation channels. 

For each of the (translationally invariant) processes identified before, we extract from this result the amplitude which approximates the thermodynamic limit best. To continue the above example, the chemical potential on the inner dimers is closest to the thermodynamic limit intuitively as well as from a graph theoretical standpoint, so we choose it as an approximation of the overall chemical potential $\mu$ in Eq.\ \eqref{eq:heffsup}. Analogous choices can be made for the other processes, though in certain cases there may be (up to symmetries) only one or even no way to embed its graph into the cluster. In the latter case, this specific process is neglected in the effective model.

Consistency between the processes and amplitudes obtained by pCUTs and gCUTs is ensured by the fact that the same quasi-particle generator is used in both cases, so both methods ultimately converge to the same effective model using different truncation schemes (orders of $J'/J$ as compared to the locality of the processes).

After approximations of all the relevant amplitudes have been obtained, the resulting, \emph{translationally invariant} effective Hamiltonian is diagonalized on the \emph{same} cluster consisting of $8$ dimers. The most relevant properties of this solution are the triplon densities (cf. Fig. \ref{fig:cluster8}) as well as the ground-state energy. Both can now be used to construct any crystal structure as explained in the main text.

\section{Self-consistent Cluster Mean Field Approximation}

In this section, we explain how the effective model in terms of pinwheel densities as given in Eq.\ \eqref{eq:pwham} can be improved using a cluster mean field approach. 

Let $\mathcal{P} = \{P_1, P_2, \dots \}$ denote the set of pinwheel structures in a certain crystal structure. The full crystal Hamiltonian which only neglects kinetic processes between dimers of different pinwheel structures then reads
\begin{equation}
 \mathcal{H}_{\rm c}= \sum_{i} \mathcal{H}_{\text{eff}}^{(i)} + \frac{1}{2} \sum_{\substack{i,j\\i\neq j}} \sum_{\substack{k\in P_i\\ l \in P_j}} V_{kl}\, \hat{n}_k \hat{n}_l 
\label{eq:crystham}
\end{equation}
where $\mathcal{H}_{\text{eff}}^{(i)}$ is the local effective Hamiltonian of pinwheel $P_i$ as in Eq.\ \eqref{eq:heffsup}.

The approximation presented in the main text basically considers the solution of the two triplon problem independently from the density-density interactions between dimers belonging to different pinwheel structures, which results in a classical Hamilton function with parametrical dependence on the pinwheel positions.  As a trivial consequence, 
the pinwheel densities as well as its chemical potential $\mu_{\rm pw}$ remain the same on any crystal structure.

In order to account for the influence of the crystal structure on the solution of the local two triplon problems, it is possible to employ a cluster mean field approach by approximating
\begin{equation}
V_{kl}\, \hat{n}_k \hat{n}_l \approx\frac{1}{2}  V_{kl}\, \left( \braket{\hat{n}_k} \hat{n}_l + \hat{n}_k \braket{\hat{n}_l} \right).
\end{equation}
We define
\begin{equation}
\mathcal{H}_{\text{eff}}^{(i)} = \mu\sum_{j \in P_i} \hat{n}_j + \mathcal{H}_{\text{eff, rest}}^{(i)}
\end{equation}
and rewrite Eq. \eqref{eq:crystham}
\begin{equation}
 \mathcal{H}_{\rm c} \approx \sum_{i} \left(  \sum_{j \in P_i} \tilde{\mu}_j \hat{n}_j + \mathcal{H}_{\text{eff, rest}}^{(i)} \right)
\end{equation}
where we introduced the effective chemical potential
\begin{equation}
\tilde{\mu}_i = \mu + \frac{1}{2} \sum_{\substack{k\\k \neq i}} \sum_{j \in \mathcal{P}_k} {V_{ij}} \braket{\hat{n}_j}
\end{equation}
which absorbs the diagonal effects of the neighbouring pinwheel structures.

This remaining model can be solved within the crystal structure's unit cell in a self consistent fashion, starting with the solution of the two triplon problem as discussed in the previous section. The relative change in the ground-state energies of the different pinwheel structures from one iteration to the next can be used as a convergence criterion. Convergence to a relative change of $10^{-9}$ is usually achieved within $10$ to $15$ iterations. 

As one would expect, the corrections obtained by this improvement grow with the density of the analysed crystal structure; for the $1/8$ structure, they are completely negligible. The fact that this extended model yields only minor (quantitative) improvements over the classical model is a strong indication that the underlying physical understanding concerning the entanglement structure of the crystals is already quite accurate.


\begin{thebibliography}{38}
\bibitem{Footnote1}
We stress that conventional plateaus made of single triplons as elementary objects have always a higher energy and are therefore not realized in the phase diagram.
\bibitem{Footnote2}
In contrast to all other plateaus, the tiny plateau at density 1/7 is only realized in the self-consistent calculation.
\end{thebibliography}
\end{document}